# Band gap renormalization and indirect optical absorption in MgSiN$_2$ at finite temperature

Dangqi Fang[1]*

[1]MOE Key Laboratory for Nonequilibrium Synthesis and Modulation of Condensed Matter, School of Physics, Xi'an Jiaotong University, Xi'an 710049, China

Abstract

We investigate the temperature effect on the electronic band structure and optical absorption property of wide-band-gap ternary nitride MgSiN$_2$ using first-principles calculations. We find that electron-phonon coupling leads to a sizable reduction in the indirect gap of MgSiN$_2$, which is indispensable in understanding the optoelectronic properties of this material. Taking the band gap renormalization into account, the band gap of MgSiN$_2$ determined by the quasiparticle $GW_0$ calculations shows good agreement with recent experimental result. The predicted phonon-assisted indirect optical absorption spectra show that with increasing temperature the absorption onset undergoes a red-shift. Our work provides helpful insights to the nature of the band gap of MgSiN$_2$ and facilitates its application in ultraviolet optoelectronic devices.

*fangdqphy@xjtu.edu.cn

# I. Introduction

The ternary heterovalent II-IV nitrides as analogs of the binary group-III nitrides have attracted a great deal of attention due to potential applications in, e.g. solar cells [1] and optoelectronic devices [2,3]. Their orthorhombic crystal structures are closely related to the wurtzite crystal structures of group-III nitrides. Since these materials consist of two cations of different valences, there is more room to tune their electronic properties. It has been found experimentally that the band gaps of $ZnSnN_2$ and $MgSnN_2$ can be tuned by controlling amounts of disorder in the cation sublattice [4–6]. The gap of group-II-IV nitrides can also be tuned by alloying. The $ZnSn_{1-x}Ge_xN_2$ films for $0 \leq x \leq 1$ have been synthesized without phase separation [7,8], the band gaps of which vary monotonically from 2.1 eV to 3.1 eV [7]. The defect physics in group-II-IV nitrides is also found to be different from the group-III nitrides [9,10]. Thus, the group-II-IV nitrides offer significantly new flexibility for band gap engineering and defect engineering, facilitating the design of optoelectronic devices.

Among the family of group-II-IV nitrides, $MgSiN_2$ powder samples have already been synthesized and shown to have an orthorhombic crystal structure (space group $Pna2_1$, no.33) [11–13]. The local bonding arrangement comprises distorted tetrahedrons so as to accommodate the two types of cations with different ionic radius [11]. $MgSiN_2$ possesses small thermal expansion [11], reasonable thermal conductivity of around 23 W/mK [14], good fracture toughness and hardness, and good oxidation resistance up to 920 °C [15], noted as a promising substrate material. Its mechanical and thermodynamic properties have been studied recently using density functional theory (DFT) calculations [16–20], showing good agreement with experimental results.

As for the band gap of MgSiN$_2$, early experimental study by measuring diffuse reflectance spectra showed a band gap of 4.8 eV [21], whereas a recent measurement at room temperature using soft X-ray absorption and emission spectroscopy indicated a band gap of 5.6 ± 0.2 eV [13]. The large band gap of MgSiN$_2$ make it suitable for ultraviolet (UV) optoelectronic applications. The theoretical studies agree that the fundamental band gap of MgSiN$_2$ is indirect [13,22–24], due to the lower symmetry of the orthorhombic crystal structure, in contrast to the direct band gap in the wurtzite-structure group-III nitride semiconductors. Nevertheless, the value of the indirect band gap spans a wide range from 4.15 eV to 6.08 eV [13,21–24], depending on the level of theory applied. Huang *et al.*, using the DFT calculations based on local density approximation (LDA), predicted an indirect band gap of 4.15 eV [22], but it is well known that DFT-LDA tends to underestimate the band gap of semiconductors. De Boer *et al.* using the modified Becke-Johnson exchange potential in combination with LDA-correlation (mBJLDA) [25,26] predicted an indirect band gap of 5.72 eV [13]. Other theoretical studies using Heyd-Scuseria-Ernzerhof (HSE06) screened hybrid functional calculations reported an indirect band gap of 5.58 eV [23]. Further studies using the quasiparticle self-consistent (QS) GW approach in the $0.8\Sigma$ approximation reported a value of 6.08 eV [24]. It is noteworthy that the measurements by De Boer *et al.* for the band gap of MgSiN$_2$ were performed at room temperature [13]. Thus, to compare with the experimental data, it is necessary to figure out the temperature effect on the electronic structure of $MgSiN_2$.

In this work, we investigate the band gap renormalization of $MgSiN_2$ at finite temperature and phonon-assisted optical absorption using first-principles calculations. We find that electron-phonon coupling leads to a sizable band gap reduction and accurate band gap of $MgSiN_2$ can be obtained by the quasiparticle $GW_0$ calculations after considering the band gap renormalization at finite temperature. The paper is organized as follows. In section II, we describe the methods for the electronic band structure calculations and the theoretical framework for the electron-phonon coupling and indirect optical absorption calculations. In section III, we present the band structure, phonon spectrum, and optical absorption of $MgSiN_2$. In section IV, we summarize our results.

## II. Computational details

A. Electronic band structure

Our first-principles DFT calculations are carried out using the projector augmented-wave method as implemented in the Vienna *ab initio* simulation package (VASP) [27–29]. The PBEsol exchange-correlation functional and a plane wave energy cutoff of 550 eV are used [30]. For the 16-atom orthorhombic unit cell of $MgSiN_2$, the Brillouin zone (BZ) is sampled by a $\Gamma$-centered 4×4×4 Monkhorst-Pack **k**-point grid [31]. All atoms are relaxed until the Hellmann-Feynman forces are less than 0.001 eV/Å. To overcome the band gap underestimation problem of the semilocal functionals, we perform additional HSE06 screened hybrid functional [32] and quasiparticle $G_0W_0$ and $GW_0$ [33] calculations based on the equilibrium structure determined by the PBEsol calculations. In the quasiparticle calculations, the initial wave functions are determined at the PBEsol level of theory, and 960 bands and an energy cutoff of 200 eV for the response function are employed.

B. Electron-phonon coupling from finite differences

The temperature dependence of the electronic band structure is caused by both thermal expansion and electron-phonon coupling. The thermal expansion can be evaluated within the quasi-harmonic approximation (QHA) [34]. We investigate the electron-phonon coupling using finite differences (FD) calculations. Based on the harmonic approximation, the vibrational average of an electronic eigenvalue at finite temperature $T$ is given by [35,36]

$$\varepsilon_{n\mathbf{k}}(T) = \frac{1}{Z}\sum_{\mathbf{S}} \langle \Phi_{\mathbf{S}}(\mathbf{u}) | \varepsilon_{n\mathbf{k}}(\mathbf{u}) | \Phi_{\mathbf{S}}(\mathbf{u}) \rangle e^{-E_{\mathbf{S}}/k_B T}, \qquad (1)$$

where $Z = \sum_{\mathbf{S}} e^{-E_{\mathbf{S}}/k_B T}$ is the partition function, $\Phi_{\mathbf{S}}(\mathbf{u})$ and $E_{\mathbf{S}}$ are the harmonic vibrational wave function and energy in state $\mathbf{S}$, $\mathbf{u} = \{u_{\mathbf{q}\nu}\}$ is a collective coordinate for all normal modes of vibrations, and $k_B$ is the Boltzmann's constant. The VASP code in conjunction with nondiagonal supercells is used to calculate the harmonic phonons [37]. Eq. (1) can be evaluated using the Monte Carlo (MC) integration technique [36]. The MC calculations, however, are computationally expensive due to the need to use large diagonal supercells to sample the vibrational BZ. The quadratic approximation and nondiagonal supercells can be used to reduce the computational cost [36,37]. The renormalization of the electronic band structure within the quadratic approximation is [38]

$$\Delta\varepsilon_{n\mathbf{k}}(T) = \frac{1}{N_\mathbf{q}} \sum_{\mathbf{q},\nu} \frac{a^{(2)}_{\mathbf{q}\nu;\mathbf{q}\nu}}{\omega_{\mathbf{q}\nu}} \left[\frac{1}{2} + n_B(\omega_{\mathbf{q}\nu}, T)\right], \quad (2)$$

where $\mathbf{q}$ and $\nu$ are the phonon indices, $N_\mathbf{q}$ is the number of the q points in the vibrational BZ, $a^{(2)}_{\mathbf{q}\nu;\mathbf{q}\nu}$ is the second-order electron-phonon coupling constant, $\omega_{\mathbf{q}\nu}$ is the phonon frequency, and $n_B$ is the Bose-Einstein population of the phonons.

C. Electron-phonon coupling from density functional perturbation theory

To provide more insights and cross-check our calculations, we use density functional perturbation theory (DFPT) to compute the phonons and associated potential derivatives, and we incorporate the electron-phonon interaction via many-body perturbation theory [35]. To lowest order in perturbation theory, the electron self-energy $\Sigma^{e-ph}$ arising from electron-phonon interactions consists of two terms, namely the Fan-Migdal (FM) and Debye-Waller (DW) terms [35,39]:

$$\Sigma^{e-ph}_{n\mathbf{k}}(\omega, T) = \Sigma^{FM}_{n\mathbf{k}}(\omega, T) + \Sigma^{DW}_{n\mathbf{k}}(T). \quad (3)$$

The frequency-dependent FM term are defined by

$$\Sigma^{FM}_{n\mathbf{k}}(\omega, T) = \sum_{m,\nu} \int_{BZ} \frac{d\mathbf{q}}{\Omega_{BZ}} |g_{mn\nu}(\mathbf{k}, \mathbf{q})|^2$$
$$\times \left[\frac{n_{\mathbf{q}\nu}(T) + f_{m\mathbf{k}+\mathbf{q}}(T)}{\omega - \varepsilon_{m\mathbf{k}+\mathbf{q}} + \omega_{\mathbf{q}\nu} + i\eta}\right.$$
$$\left. + \frac{n_{\mathbf{q}\nu}(T) + 1 - f_{m\mathbf{k}+\mathbf{q}}(T)}{\omega - \varepsilon_{m\mathbf{k}+\mathbf{q}} - \omega_{\mathbf{q}\nu} + i\eta}\right], \quad (4)$$

where $f_{m\mathbf{k}+\mathbf{q}}(T)$ and $n_{\mathbf{q}\nu}(T)$ are the Fermi-Dirac and Bose-Einstein occupation functions. The electron-phonon matrix elements $g_{mn\nu}(\mathbf{k}, \mathbf{q})$ are given as

$$g_{mn\nu}(\mathbf{k},\mathbf{q}) = \langle \psi_{m\mathbf{k}+\mathbf{q}} | \Delta_{\mathbf{q}\nu} V^{KS} | \psi_{n\mathbf{k}} \rangle, \tag{5}$$

with $\Delta_{\mathbf{q}\nu} V^{KS}$ the first-order variation of the self-consistent KS potential induced by the phonon mode $\mathbf{q}\nu$.

The static DW term depends on the second-order derivative of the KS potential and is approximated as [35,39]

$$\Sigma_{n\mathbf{k}}^{DW}(T) = \sum_{\mathbf{q}\nu m} \left[ 2n_{\mathbf{q}\nu}(T) + 1 \right] \frac{g_{mn\nu}^{2,DW}(\mathbf{k},\mathbf{q})}{\varepsilon_{n\mathbf{k}} - \varepsilon_{m\mathbf{k}}}, \tag{6}$$

where $g_{mn\nu}^{2,DW}(\mathbf{k},\mathbf{q})$ is an effective matrix element that, within the rigid-ion approximation, can be expressed in terms of the $g_{mn\nu}(\mathbf{k},\mathbf{q})$ matrix elements.

In the non-adiabatic Allen-Heine-Cardona (AHC) approach [40,41], the zero-point and temperature-dependent renormalizations of the band structure are obtained from the real part of the self-energy using the on-the-mass-shell approximation

$$\Delta \varepsilon_{n\mathbf{k}}^{AHC}(T) = \text{Re}\left[ \Sigma_{n\mathbf{k}}^{e-ph}(\varepsilon_{n\mathbf{k}}^{0}, T) \right], \tag{7}$$

where $\varepsilon_{n\mathbf{k}}^{0}$ is the bare DFT eigenvalue.

Calculations are performed with the ABINIT (version 9.4.0) code [42,43] using the PBEsol functional [30] and PseudoDojo norm-conserving pseudopotentials [44]. The plane wave kinetic energy cutoff is set to be 40 Ha. The BZ sampling of 4×4×4 Γ-centered **k**-point grid is used to obtain the electronic ground state density of the unit cell. A coarse 4×4×4 Γ-centered **q**-point grid is employed in the phonon calculations using DFPT. In the electron-phonon calculations, in order to accelerate the convergence with the number of empty bands, the contributions above band index 60 are replaced with the solution of a non-self-consistent Sternheimer equation [45], where 12 empty bands are used. We interpolate the DFPT potential and self-energy onto a 28×28×28 Γ-centered **q**-point grid, which converges the renormalization within a few meV.

D. Indirect optical absorption

The frequency-dependent dielectric matrix in the long-wavelength limit and the optical properties are calculated at the independent particle level using the PBEsol functional and VASP code. The imaginary part of dielectric function is given as [46]

$$\varepsilon_2^{\alpha\beta}(\omega) = \frac{4\pi^2 e^2}{\Omega} \lim_{q \to 0} \frac{1}{q^2} \sum_{c,v,\mathbf{k}} 2w_{\mathbf{k}} \delta(\varepsilon_{c\mathbf{k}} - \varepsilon_{v\mathbf{k}} - \omega) \\ \times \langle u_{c\mathbf{k}+\mathbf{e}_\alpha q}|u_{v\mathbf{k}}\rangle \langle u_{c\mathbf{k}+\mathbf{e}_\beta q}|u_{v\mathbf{k}}\rangle^* \qquad (8)$$

where the indices $c$ and $v$ refer to the conduction band and valence band states respectively, the vectors $\mathbf{e}_\alpha$ are unit vectors for the three Cartesian directions, and $u_{c\mathbf{k}}$ is the cell periodic part of the orbitals at the **k** point. A Gaussian function with a width of 50 meV is substituted for the delta function in Eq. (8) that ensures energy conservation during the absorption of a photon.

The finite temperature dielectric matrix is calculated using

$$\varepsilon_2(\omega, T) = \frac{1}{Z} \sum_{\mathbf{S}} \langle \Phi_\mathbf{S}(\mathbf{u})|\varepsilon_2(\omega, \mathbf{u})|\Phi_\mathbf{S}(\mathbf{u})\rangle e^{-E_\mathbf{S}/k_B T}, \qquad (9)$$

as originally proposed by Williams [47] and Lax [48], which takes the indirect transition into account by including the phonons in the supercell. We reduce the plane wave energy cutoff to 350 eV (without loss of accuracy) and we use 2×2×2 supercell of MgSiN$_2$ containing 128 atoms, together with a 4×4×4 Γ-centered **k**-point grid. We evaluate the corresponding integral using the MC technique in conjunction with thermal lines method to accelerate the sampling [49]. Converged results are obtained using four sampling points.

In order to address the problem of band gap underestimation we apply a scissor correction $\Delta$ chosen so as to reproduce the $GW_0$ band gap. The imaginary parts with scissor are defined as $\varepsilon_{2,s}(\omega) = (1-\Delta/\hbar\omega)\varepsilon_2(\omega-\Delta/\hbar)$ to fulfill the $f$-sum rule [50,51]. The corresponding real parts $\varepsilon_{1,s}$ are obtained using the Kramers-Kronig relation, and the absorption coefficients are calculated as

$$\alpha(\omega) = \frac{\sqrt{2}\omega}{c}[\sqrt{\varepsilon_{1,s}^2 + \varepsilon_{2,s}^2} - \varepsilon_{1,s}]^{\frac{1}{2}}, \qquad (10)$$

where $c$ is the speed of light.

### III. Results and Discussion

A. Structural and electronic properties

The orthorhombic crystal structure of MgSiN$_2$ is shown in the top panel of Fig. 1. In order to satisfy local charge neutrality, each of N atoms has two Mg and two Si atoms as the nearest neighbors. The optimized lattice constants are $a$=5.276 Å, $b$=6.470 Å, and $c$=4.994 Å, which agree well with previous theoretical and experimental results as shown in Table I. The calculated band structure is shown in the bottom panel of Fig.1. The lowest conduction band is mainly derived from the $s$ states of N, Mg, and Si atoms, giving rise to a large band dispersion. The electron effective masses along the Γ-X, Γ-Y, and Γ-Z directions are 0.36, 0.35, and 0.36 (in units of free electron mass $m_e$), respectively. The valence bands between 0 eV and -6 eV have predominantly N 2$p$ character and are slightly hybridized with Mg 3$p$ and Si 3$p$ states, while those between -6 eV and -8 eV have hybridization between N 2$p$ and Si 3$s$ states. The calculated PBEsol band gaps are 4.06 eV and 4.48 eV for the indirect T-Γ and direct Γ-Γ transitions, respectively. The HSE06, $G_0W_0$, and $GW_0$ calculations verify the indirect band gap nature of MgSiN$_2$, while the magnitude of the indirect gap varies from 5.52 eV to 6.11 eV depending on the level of theory applied, as shown in Table II. The energy differences between the direct and indirect gaps being 0.45 eV, 0.47 eV, and 0.48 eV in the HSE06, $G_0W_0$, and $GW_0$ calculations, respectively, are slightly larger than the value of 0.42 eV in the PBEsol calculations. In Table II, we also show the band gaps obtained in earlier computational studies, and our calculations are in good agreement with these previous studies. We note that our calculated $GW_0$ band gaps are similar to the 0.8Σ-QSGW ones in which an empirical scale factor 0.8 is used.

Recent experiments report the band gap of MgSiN$_2$ is 5.6±0.2 eV evaluated using soft X-ray absorption and emission spectroscopy [13]. It is noteworthy that the experimental measurements for the band gap are performed at room temperature, while all theoretical calculations are based on equilibrium-static lattice. Therefore, it is necessary to take the temperature effect into account when comparing the theoretical results with the experimental data.

B. Lattice dynamics

The phonon dispersions of MgSiN$_2$ along the high-symmetry lines of the BZ are shown in Fig.2, which are calculated using the FD method in conjunction with nondiagonal supercells [37]. The vibrational BZ is sampled using a coarse 4×4×4 **q**-point grid and the phonon dispersions are then constructed by the Fourier interpolation to a finer grid. The LO-TO splitting is incorporated by calculating the dielectric constants and Born effective charges and using the scheme described in Ref. [52]. The discontinuity at zone center induced by LO-TO splitting is observed, where the modes with higher frequencies have stronger effect, which is in good agreement with previous calculations [17,18]. Moreover, we find a frequency gap of about 30 cm$^{-1}$ in the vibrations above 800 cm$^{-1}$. Our calculated phonon frequencies show good agreement with the experimental Raman frequencies and the previous theoretical results [17,18].

C. Band gap renormalization

Figure 3 (a) shows the temperature dependence of the indirect band gap renormalization of MgSiN$_2$ calculated using the PBEsol functional and the FD scheme. In the quadratic approximation calculations, we use **q**-point grids of different sizes and finite displacement amplitude of $0.5\sqrt{\langle u_{v\mathbf{q}}^2 \rangle}$. Thanks to the nondiagonal supercell method [37], we can sample the vibrational BZ up to 5×5×5 **q**-point grid using FD with the largest nondiagonal supercell containing 80 atoms. Our results indicate that electron-phonon coupling leads to a large reduction in the indirect band gap of MgSiN$_2$ and the magnitude of band gap renormalization increases with increasing temperature. The convergence of band gap renormalization with the size of **q**-point grid is somewhat slow. The zero-point renormalization (ZPR) is -0.311 eV calculated using 5×5×5 **q**-point grid and the result corresponding to 4×4×4 **q**-point grid is within 0.025 eV of that calculated using 5×5×5 **q**-point grid. We also calculate the indirect band gap renormalization of MgSiN$_2$ using the non-adiabatic AHC method from DFPT as shown in Fig. 3 (b). Using the interpolation technique with a 28×28×28 **q**-point grid, the ZPR from the AHC calculation is converged to be -0.363 eV, the absolute value of which is larger than that of -0.238 eV estimated empirically in Ref. [24]. The two computational methods based on the quadratic approximation, FD and AHC, show a similar trend in the change of band gap renormalization with temperature, validating the band gap reduction due to electron-phonon coupling for MgSiN$_2$. We note that for the ZPR calculated using the **q**-point grid of 4×4×4, the AHC result is smaller by 14% in magnitude than the FD quadratic approximation calculation, which probably stems from the non-adiabatic effect and the lack of the off-diagonal components of DW self-energy [53] in the AHC method used. Miglio *et al.* computed the ZPRs of 30 materials using first-principles non-adiabatic AHC method, showing that for materials with light elements the absolute value of band gap renormalization is often larger than 0.3 eV, and up to 0.7 eV (e.g., -399 meV for AlN, -524 eV for MgO, and -699 meV for BeO) [54]. Therefore, it is necessary to take the ZPRs into account for accurately predicting the band gaps of these materials. Our calculated ZPR for MgSiN$_2$ using the non-adiabatic AHC method is comparable to that for AlN obtained by Miglio *et al.* [54], mainly because of their analogous crystal structures and constituent light elements.

To elucidate the mechanism of the ZPR, we examine the individual phonon contributions to the renormalization. Rearranging the Eq. (2), we can write

$$\Delta\varepsilon_{n\mathbf{k}}(T) = \sum_{\mathbf{q},\nu} \frac{1}{N} \frac{a^{(2)}_{\mathbf{q}\nu;\mathbf{q}\nu}}{\omega_{\mathbf{q}\nu}} \left[\frac{1}{2} + n_B(\omega_{\mathbf{q}\nu}, T)\right] = \sum_{\mathbf{q},\nu} \Delta\varepsilon_{n\mathbf{k},\mathbf{q}\nu}(T) \qquad (11)$$

to obtain the contribution from each phonon, where $N$ is the number of the q points in the vibrational BZ. For this analysis, we use a 5×5×5 **q**-point grid. In Fig. 4, we plot the renormalization of CBM and VBM induced by different phonon modes. High-frequency optical phonon modes are found to have the largest contributions. We further employ the generalized Fröhlich model [54] and perform calculations at **q** = Γ as implemented in the ABINIT. Converged results are obtained using the plane wave kinetic energy cutoff of 40 Ha and Γ-centered **k**-point grid of 4×4×4. The ZPR for MgSiN$_2$ from the generalized Fröhlich model is found to be -0.253 eV. These results indicate that the large bandgap renormalization in MgSiN$_2$ is dominated by the Fröhlich mechanism.

The quadratic approximation methods rely on a low-order expansion of the electronic state as a function of the atomic displacements, which only treat electron-phonon interaction including single-phonon process. To examine the effect of higher-order terms, we use MC integration technique with 100 different atomic configurations for sampling to compute the band gap renormalization. The MC calculations require the use of diagonal supercell and thus is computationally demanding. The results for the MC calculations using 2×2×2, 3×3×3, and 4×4×4 supercells are displayed in Fig. 3 (a). The magnitude of band gap renormalization become larger with increasing the size of supercell, but the convergence is slow. The 4×4×4 supercell of MgSiN$_2$ contains 1024 atoms, which is the largest system we can afford for the MC calculations. Here, we analyze the band gap renormalizations based on the MC calculations using the 4×4×4 supercells. The zero-point and 300 K indirect band gap renormalizations are -0.393 ± 0.006 eV and -0.520 ± 0.007 eV, respectively, which are 8% and 17% larger in magnitude than the corresponding converged values in the AHC calculations. Moreover, it can be seen from Fig. 3 that the slope of the temperature-dependent band gap renormalization for the MC calculations is steeper than that for the AHC and FD quadratic approximation calculations.

The non-adiabatic AHC method and the MC technique are two of the state-of-the-art approaches to study the electron-phonon interaction, which have been recently reviewed [35,36]. Furthermore, J.-M. Lihm and C.-H. Park generalized the AHC theory to cope with the off-diagonal components of DW self-energy and the electron-phonon coupling induced wave-function hybridization [53]. M. Zacharias and F. Giustino proposed a special displacement method (SDM) to compute the electronic and optical properties of solids at finite temperature using a distorted single supercell [55,56]. It is worth noting that both the MC technique and the SDM are based on the adiabatic approximation, which miss the non-adiabatic effects. Recently, Miglio *et al.* highlighted the importance of non-adiabatic effects in ZPR of the electronic band gap and argued that for infrared-active materials the adiabatic supercell (ASC) method avoids the divergence problem for wrong reason by sampling a whole set of amplitudes [54]. The ZPRs of the fundamental gaps of several materials are listed in Table III. The one-shot ASC calculations in Ref. [57] give good ZPR estimations for crystalline C and Si in the diamond structure, but underestimate the ZPR for polar semiconductor GaN. The results obtained from the non-adiabatic AHC calculations totally agree well with available experimental data.

The electron-electron correlation may affect the electron-phonon coupling strength. Antonius *et al.* found that many-body effects modify the electron-phonon coupling energies by more than 40% in the direct gap ZPR of diamond [58]. Nevertheless, Karsai *et al.* calculated the ZPRs of 18 semiconductors and found that for the majority of investigated compounds the $G_0W_0$ corrections to the renormalization are very small [57]. We defer the investigation of the many-body effect on the band gap renormalization of $MgSiN_2$ to future work.

The thermal expansion of $MgSiN_2$ have been investigated by Råsander *et al.* using DFT in combination with the quasi-harmonic approximation [19]. Our calculated equilibrium-static lattice constants are in good agreement with that in Ref. [19], as shown in Table I. Using the lattice constants at 0 K and 300 K in Ref. [19] calculated using the PBEsol functional, we obtain the indirect band gap renormalization of $MgSiN_2$ arising from the thermal expansion to be -0.085 eV at 0 K due to zero-point phonon effects and -0.099 eV at 300 K. The band gap reduction induced by thermal expansion is much smaller than that by electron-phonon coupling.

We list the results of the indirect band gap renormalization for $MgSiN_2$ in Table IV. Taking thermal expansion contribution into account, we obtain the indirect band gap correction to be about -0.48 eV (-0.45 eV) at 0 K and -0.62 eV (-0.54 eV) at 300 K from the MC (AHC) calculations. For $MgSiN_2$, comparing with recent experimental data of 5.6 ± 0.2 eV measured at room temperature [13], the $GW_0$ result agrees well with the experimental value only when the band gap renormalization at 300 K is taken into account.

D. Phonon-assisted optical absorption

To investigate the temperature effect on the optical absorption spectrum of $MgSiN_2$, we calculate its frequency-dependent absorption coefficient at the equilibrium-static lattice, 0 K, and 300 K. Figure 5 (a) presents the *xx*, *yy*, and *zz* components of frequency-dependent absorption coefficient of $MgSiN_2$. A scissor correction of 2.05 eV is employed so that the indirect band gap is corrected to the $GW_0$ value. The *zz* component of absorption spectrum based on the equilibrium-static lattice exhibits an absorption peak around 6.5 eV, corresponding to the direct transition at the Γ point. Nonzero values below the direct gap result from Gaussian broadening we use. To assess the indirect absorption, phonon-assisted transitions need to be considered, which is effectively accomplished in our calculations. As shown in Fig. 5 (a), our calculated spectra at 0 K and 300 K correctly capture indirect absorptions. The Tauc plot in Fig. 5 (b) verifies the occurrence of indirect optical absorption. The absorption onset at 0 K is governed by zero-point quantum motion, which enables the indirect transition to occur. Increasing temperature leads to a red-shift of the absorption onset, consistent with the findings of temperature-dependent band gap shown in Fig. 3.

## IV. Conclusions

In this work, we have investigated the band gap renormalization and optical absorption property of $MgSiN_2$ at finite temperature using first-principles calculations. Electron-phonon coupling reduces the indirect band gap of $MgSiN_2$ significantly and enables phonon-assisted optical absorption process across the minimum indirect gap of $MgSiN_2$ to occur. High-frequency optical phonon modes have the largest contributions to the zero-point renormalization. The large band gap renormalization in $MgSiN_2$ is dominated by the Fröhlich mechanism. Our calculations demonstrate that for $MgSiN_2$ good agreement with the experimental band gap is obtained by the quasiparticle $GW_0$ calculations in conjunction with finite-temperature band gap renormalization. The predicted indirect absorption spectra show that with increasing temperature the absorption onset undergoes a red-shift. Our work will facilitate the application of $MgSiN_2$ in ultraviolet optoelectronic devices. The influences of the electron-electron correlation and the off-diagonal components of the Debye-Waller self-energy on the band gap renormalization are worth further investigating.

**Supplementary Material**

See supplementary material for the theoretical details and for the electronic band structure, phonon spectrum, and convergence test for $MgSiN_2$.

**ACKNOWLEDGMENTS**

We gratefully acknowledge B. Monserrat for the helpful discussions. We acknowledge the financial support from the National Natural Science Foundation of China (Grant No. 11604254), the Natural Science Foundation of Shaanxi Province (Grant No. 2019JQ-240), and the China Scholarship Council (No. 201906285031). We also acknowledge the HPCC Platform of Xi'an Jiaotong University for providing the computing facilities.

**Conflict of interest**

The authors have no conflicts to disclose.

**Data Availability Statement**

The data that support the findings of this study are available from the corresponding author upon reasonable request.

**Figure captions**

FIG.1 Top panel: Top and side views of the 16-atom primitive cell and Brillouin zone of the orthorhombic crystal structure (space group Pna2$_1$) for MgSiN$_2$. Mg, Si, and N atoms are represented by bronze, blue, and grey balls, respectively. High symmetry points are marked with labels. Bottom panel: Electronic band structure of MgSiN$_2$ calculated using the PBEsol functional. The top of valence band is set to zero.

FIG. 2 Phonon dispersion of MgSiN$_2$ calculated using a coarse 4×4×4 **q**-point grid including the LO-TO splitting effect. Experimental Raman frequencies are shown using the red circles [17].

FIG. 3 Temperature dependence of the indirect band gap renormalization of MgSiN$_2$ resulting from electron-phonon coupling obtained by the quadratic approximation (QA) (solid lines) and MC calculations (filled points) from finite differences (a) and the non-adiabatic AHC calculations from density functional perturbation theory (b). The statistical error bars are included in the MC data points.

Fig. 4 Renormalization of the CBM and VBM induced by different phonon modes (total number is 2160).

FIG. 5 (a) *xx*, *yy*, and *zz* components of the absorption coefficient for MgSiN$_2$ based on the PBEsol calculations with a scissor correction. Calculations correspond to the stationary atoms at the equilibrium positions (dashed lines), and to the phonon-assisted contributions at 0 K (dotted lines) and 300 K (solid lines). 2×2×2 supercells, 4×4×4 Γ-centered **k**-point grid, and a Gaussian broadening of 50 meV are used for the calculations. (b) Tauc plot for the indirect absorption in MgSiN$_2$ at 300 K.

TABLE I. Calculated and experimental lattice constants of MgSiN$_2$. The lattice constants in Ref. [19] are calculated using the PBEsol functional.

|  | $a$ (Å) | $b$ (Å) | $c$ (Å) |
| --- | --- | --- | --- |
| This work | 5.276 | 6.470 | 4.994 |
| Ref. [19] (static) | 5.2768 | 6.4694 | 4.9930 |
| Ref. [19] (T=0 K) | 5.2945 | 6.4861 | 5.0108 |
| Ref. [19] (T=300 K) | 5.2973 | 6.4888 | 5.0137 |
| Expt. (T=10 K) [11] | 5.27078(5) | 6.46916(7) | 4.98401(5) |
| Expt. (T=300 K) [11] | 5.27249(4) | 6.47334(6) | 4.98622(4) |

TABLE II. Calculated band gaps of MgSiN$_2$ using different methods based on the equilibrium-static lattice. QSGW represents the quasiparticle self-consistent method. 0.8Σ-QSGW is an empirical approach with ($V_{xc}^{QSGW}$ - $V_{xc}^{GGA}$) multipled by a factor 0.8 [24].

| This work | Indirect (T→Γ) | Direct (Γ→Γ) |
| --- | --- | --- |
| PBEsol | 4.06 | 4.48 |
| HSE06 | 5.52 | 5.97 |
| G$_0$W$_0$ | 5.79 | 6.26 |
| GW$_0$ | 6.11 | 6.59 |
| Previous work | | |
| LDA [22] | 4.15 | |
| GGA-PBE [24] | 4.01 | 4.44 |
| mBJLDA [13] | 5.72 | |
| HSE06 [23] | 5.58 | |
| 0.8Σ-QSGW [24] | 6.08 | 6.53 |
| Experiment [13] | 5.6 ± 0.2 | |
| Experiment [21] | 4.8 | |

TABLE III. Zero-point renormalization (ZPR) energies of the fundamental gaps of several materials arising from thermal lattice expansion (TLE) and electron-phonon coupling (EPC). Abbreviations for crystal structure: dia=diamond, w=wurtzite, rs=rocksalt. ASC denotes the adiabatic supercell method.

| Material | ZPR-exp (meV) | ZPR-TLE (meV) [54] | ZPR-EPC-AHC (meV) [54] | ZPR-EPC-ASC (meV) [57] |
|---|---|---|---|---|
| C-dia | -334 [59], -370 [60] | -27 | -330 | -320 |
| Si-dia | -72 [59], -64 [60] | +9 | -56 | -65 |
| GaN-w | -180 [59], -173 [60] | -49 | -189 | -94 |
| AlN-w | -350 [59], -239 [60], -483 [61] | -85 | -399 | |
| MgO-rs | | -117 | -524 | |

TABLE IV. Indirect band gap renormalization of MgSiN$_2$. The MC results are obtained using 4×4×4 supercell calculations and the non-adiabatic AHC results using 28×28×28 **q**-point sampling. The theoretical band gaps, E$_g$(TLE+MC$^{4\times4\times4}$) and E$_g$(TLE+AHC), are obtained from the GW$_0$ value with the band gap renormalizations incorporated. The experimental band gap E$_g$(exp) is taken from Ref. [13].

|       | TLE (meV) | EPC-MC$^{4\times4\times4}$ (meV) | EPC-AHC (meV) | E$_g$(TLE+MC$^{4\times4\times4}$) (eV) | E$_g$(TLE+AHC) (eV) | E$_g$(exp) (eV) |
|-------|-----------|----------------------------------|---------------|----------------------------------------|---------------------|-----------------|
| 0 K   | -85       | -393                             | -363          | 5.63                                   | 5.66                |                 |
| 300 K | -99       | -520                             | -444          | 5.49                                   | 5.57                | 5.6 ± 0.2       |

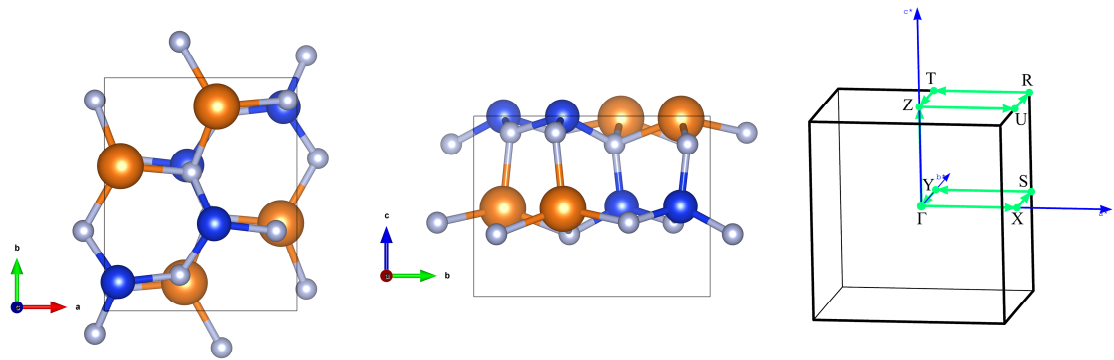

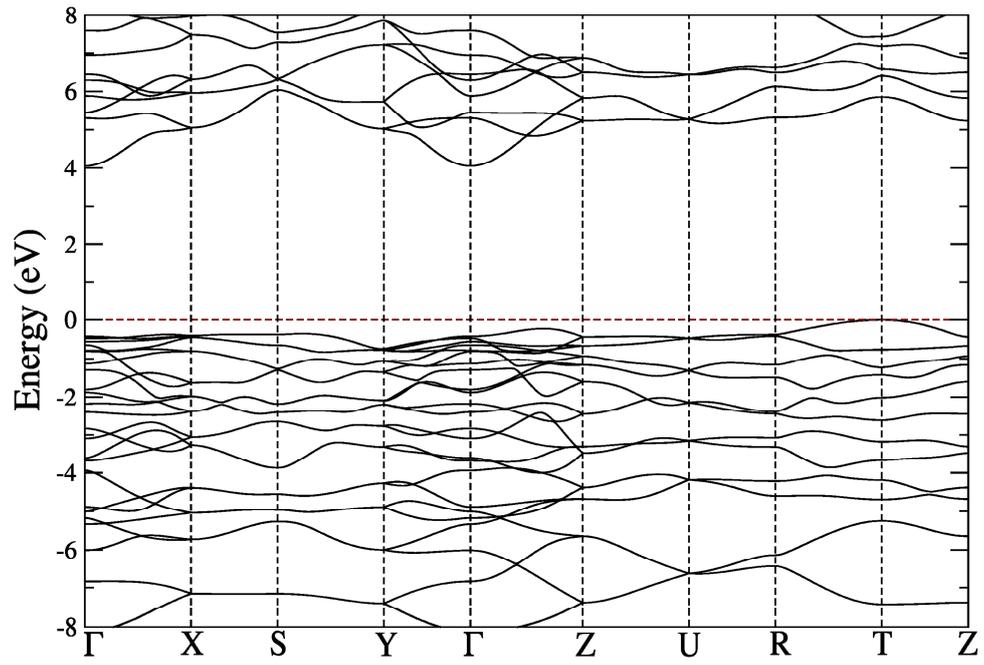

Fig. 1

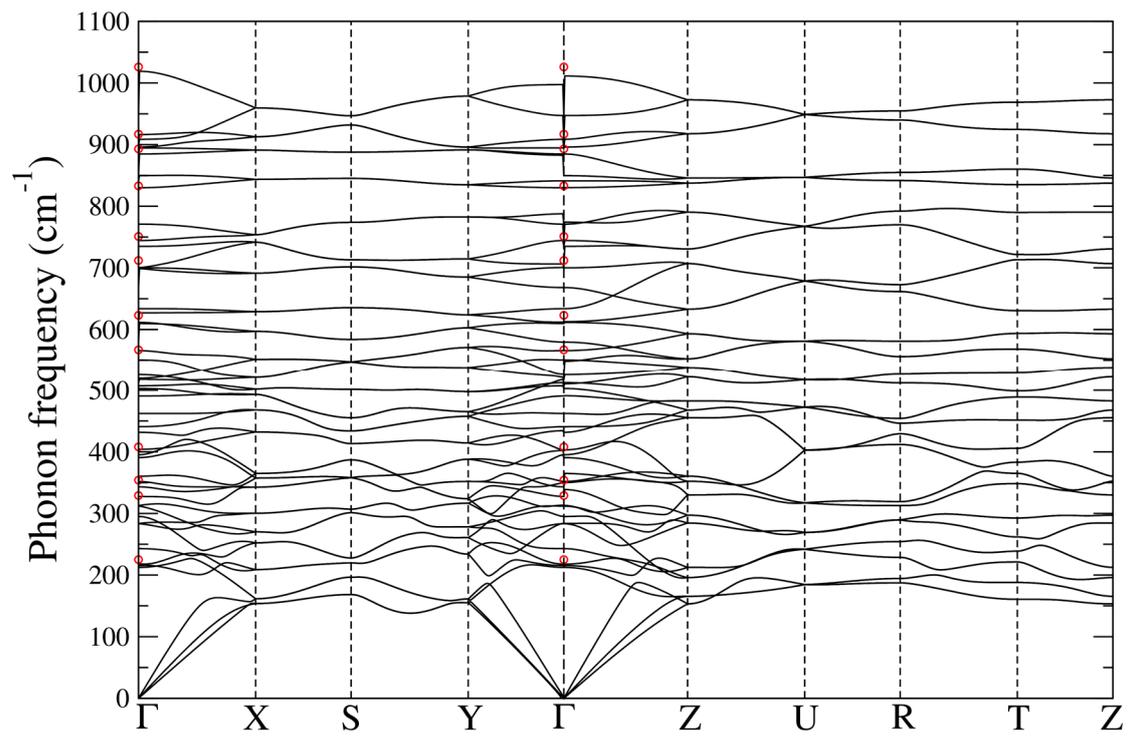

Fig. 2

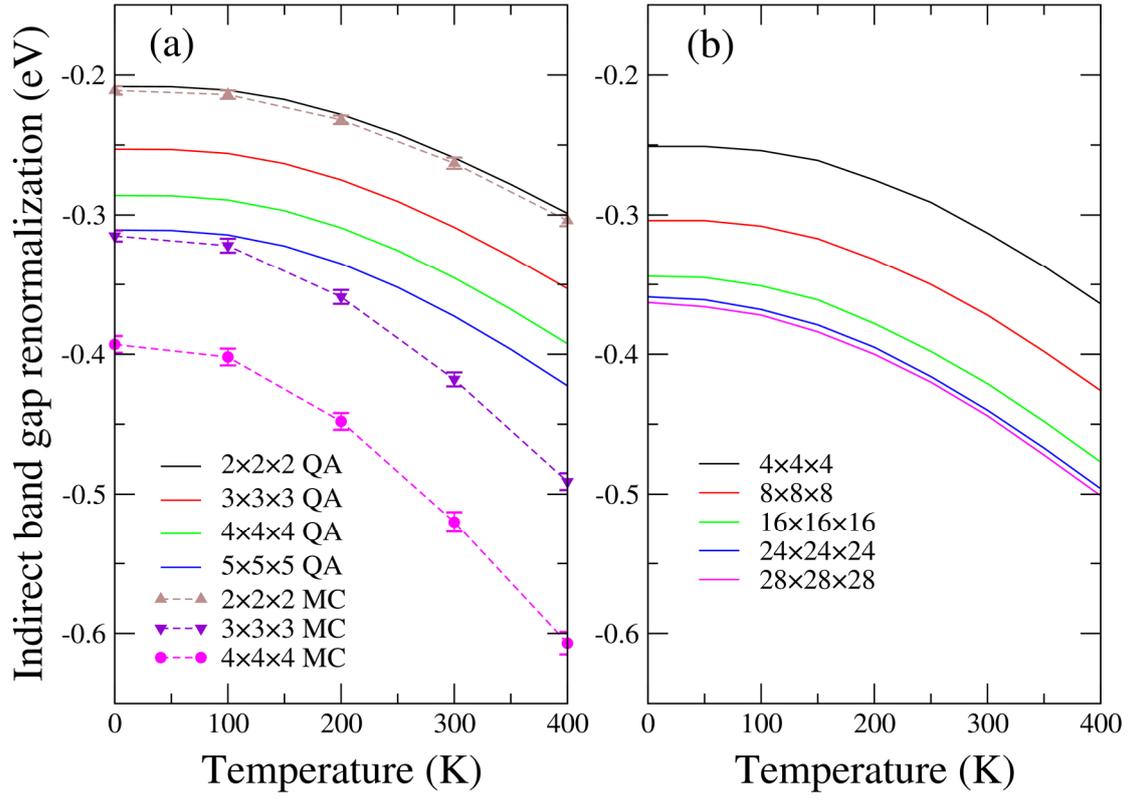

Fig. 3

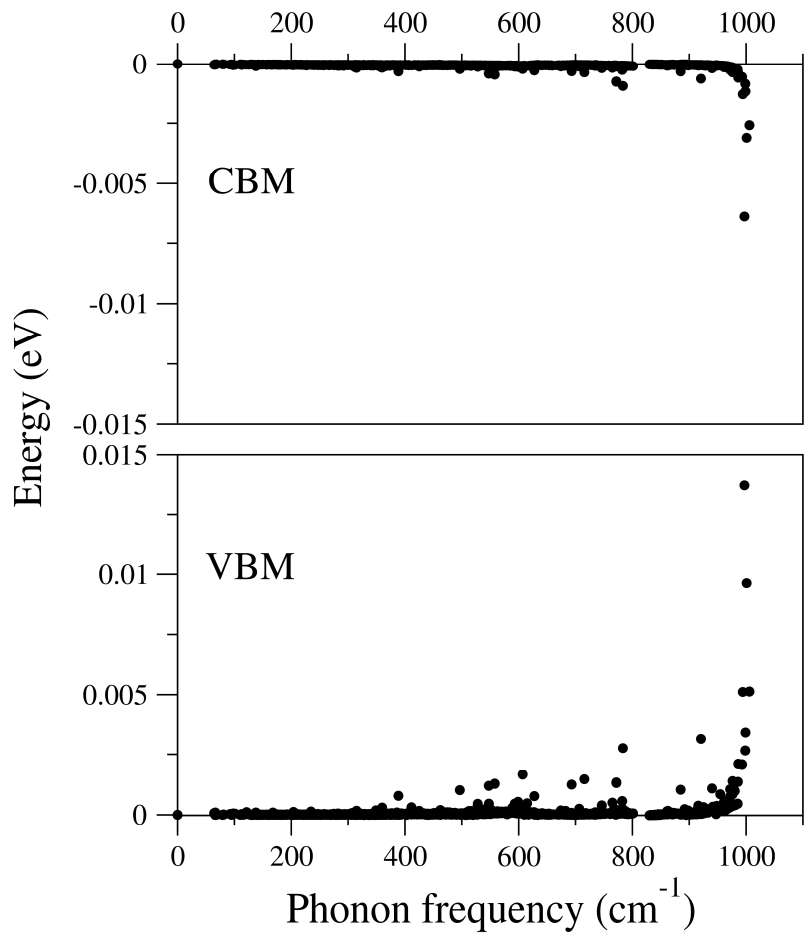

Fig. 4

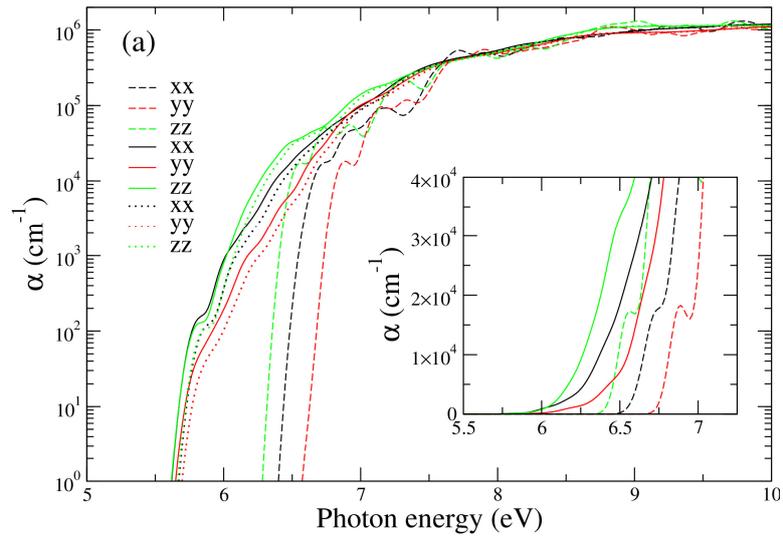

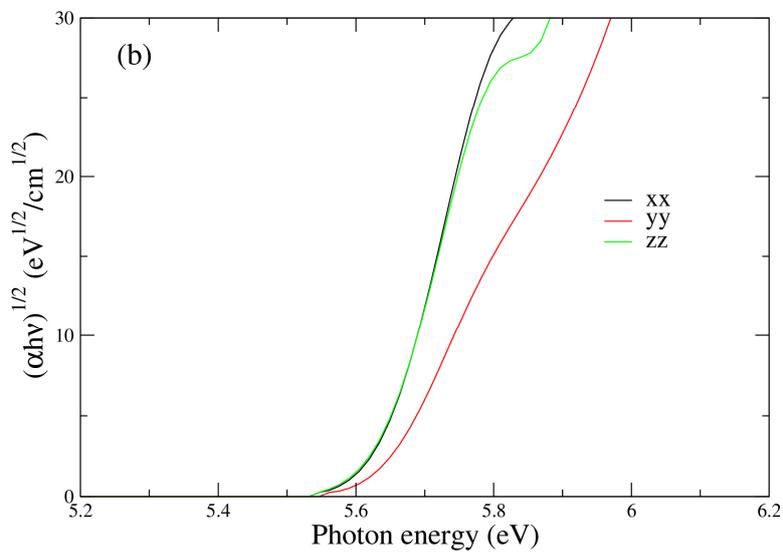

Fig. 5